\documentclass{optica-article}

\journal{opticajournal} 

\articletype{Research Article}

\usepackage{tikz}
\usetikzlibrary{trees,arrows}
\usepackage{comment}
\usepackage{algorithm}
\usepackage{algpseudocode}
\usepackage{graphicx}

\usepackage{amssymb}
\usepackage{braket}
\usepackage[utf8]{inputenc} 
\usepackage[T1]{fontenc} 
\usepackage{url}
\usepackage{ifthen}
\usepackage{cite}

\begin{document}

\title{Bayesian Greedy Receiver for Pulse Position Modulation without an Error Floor under Thermal Noise}

\author{Leo Bia,\authormark{1,*} Christos N. Gagatsos,\authormark{2,1,3} and Saikat Guha \authormark{4,1}}

\address{
\authormark{1}Wyant College of Optical Sciences, The University of Arizona, Tucson, AZ, USA\\
\authormark{2}Department of Electrical and Computer Engineering, The University of Arizona, Tucson, AZ, USA\\
\authormark{3}Program in Applied Mathematics, The University of Arizona, Tucson, Arizona 85721, USA\\
\authormark{4} Department of Electrical and Computer Engineering, University of Maryland, College Park, Maryland, 20742, USA
}

\email{\authormark{*}lclashinbia@arizona.edu} 

\begin{abstract*}
We study quantum receiver architectures for M-ary pulse-position modulation (PPM) in thermal noise and the photon-starved regime. Building on greedy decision strategies, we examine displacement-squeezing configurations and introduce a Bayesian "slicing" receiver that partitions each slot into slices and performs sequential posterior updates. Simulations for 4-PPM indicate gains over standard greedy and CPN receiver under background noise, with no apparent error floor for the sliced Bayesian scheme, the effect appears attributable to inference accumulation via Bayes updates. We discuss computational trade-offs and outline directions for characterizing the sliced Bayesian receivers and its relations to Helstrom performance.
\end{abstract*}

\section{Introduction}
The $M$-ary Pulse Position Modulation (PPM) is a technique wherein a single laser-light pulse is placed in one of $M$ time slots and vacuum in the remaining $M-1$ time slots, each $T$ seconds duration, to convey information. The pulse can be represented by a generally constant complex amplitude $E\in\mathbb{C}$ temporal waveform of time duration $T$-seconds which can also be described by the coherent state $\ket{\alpha}$, related by the mean photon number $N=|E|^2T=|\alpha|^2$.  Each PPM codeword spans $M$ slots, i.e. a total duration of $MT$ seconds. The quantum state of the $i$-th codeword is given by
\begin{equation}
\ket{\Psi_i} = \ket{0}_1\ket{0}_2\ldots\ket{\alpha}_i\ldots\ket{0}_M,
\label{eqn:PPM}
\end{equation}
where $i \in \{1, \ldots, M\}$ denotes the pulse bearing slot. The state $\ket{\Psi_i}$ represents a tensor product over $M$ orthogonal time modes, with the coherent state $\ket{\alpha}$ occupying the $i$-th slot and vacuum $\ket{0}$ elsewhere. We assume a uniform prior probability for each codeword being transmitted, i.e., $q=\text{Pr}[i \text{ is transmitted}] = 1/M$. 

Coherent states are not mutually orthogonal, i.e. $\braket{\alpha|\beta}\neq 0$ for $\alpha \neq \beta$, thus cannot be perfectly distinguished by any quantum measurement. The quantum limit on the minimum probability of error, known as the Helstrom bound~\cite{Helstrom1969}, can be evaluated for any set of coherent-state codewords using the Yuen–Kennedy–Lax (YKL) conditions~\cite{Yuen1975}. In the absence of noise, the Helstrom limit for $M$-ary PPM admits a closed-form expression~\cite{Krovi2015}:
\begin{equation}
    P_{\mathrm{e,min}} = \frac{M-1}{M^2} \left[ \sqrt{1 + (M-1)e^{-N}} - \sqrt{1 - e^{-N}} \right]^2.
    \label{eqn:helstrom_PPM}
\end{equation}
This expression captures the ultimate quantum advantage in discriminating PPM codewords using an optimal joint measurement over all $M$ slots. In the high mean photon regime $Me^{-N}\ll1$ the $P_{\mathrm{e,min}} \sim e^{-2N}$. Although the Helstrom limit can be derived for any constellation of codewords, it does not suggest a practical receiver architecture to achieve this minimum error probability.

PPM is the leading modality for long-distance optical communication due to its high special efficiency and low beam divergence. In deep-space or inter-satellite links, severe power and aperture constraints push systems into the photon-starved regime, where the average received photon number per mode satisfies $N \ll1$ \cite{boroson}. In this regime, where only a few photon per mode are detected, performances gains must rely on more efficient modulation formats, advanced receiver architectures, and powerful error-correcting codes. The ultimate capacity of such a pure-state quantum system (like the lossy bosonic channel) is given by the Holevo bound \cite{Holevo1973}, and it can, in principle, be achieved using a structured optical receiver. This receiver consists of a global unitary transformation acting on the entire quantum codeword, followed by local projective measurements on individual symbols \cite{guha_holevo,banaszek}. By enabling joint detection, this architecture can achieve superadditive capacity, outperforming conventional symbol-by-symbol receivers in the low-photon-number regime. It is precisely in this regime that PPM demonstrates its exceptional efficiency. 

One of the most notable demonstrations is NASA's Mars Lasers Communication Demonstration (MLCD) project, which adopted high-order PPM (i.e. 64-PPM) as its baseline modulation scheme to support data rates of 1-10 Mbps over a link spanning 400 million kilometers between Mars and Earth \cite{boroson2004}. Several studies have applied PPM to deep-space communication scenarios, analyzing its performance in the photon-starved regime under realistic power and noise constraints. These examine Photon Information Efficiency (PIE=$\tfrac{C}{N}$), where $C$ is the capacity in bits per mode, and capacity $C$ bounds \cite{jarzynapie,banaszekexcessnoise}, the use of high-order PPM and soft decoding to approach quantum limits \cite{jarzynasoftdecoding,zwolinskirange}, and structured receivers to enhance efficiency under limited peak power \cite{jarzynaoptimization}.

In this work, we extend the adaptive receiver framework for PPM by presenting new performance results for greedy demodulation strategies~\cite{Lukanowski:24} that incorporate both optimized displacement and phase-sensitive amplification. We analyze two new variants of the greedy receiver: one that jointly optimizes the displacement and squeezing parameters, and another that applies a time-varying displacement waveform at each time slot. Most notably, we introduce a novel algorithm—the slicing Bayesian greedy receiver—which builds upon the slicing method previously explored in~\cite{bia}. This receiver divides each PPM time slot into multiple optical slices and maintains the full posterior distribution over all codeword hypotheses to optimize the measurement strategy at each slice.

In Section II, we define the detection statistics for displaced thermal states under various unitary operations, including displacement, squeezing, and time-varying waveforms. Section III reviews fundamental binary discrimination strategies—including direct detection, optimal displacement, and the Dolinar receiver—that serve as building blocks for PPM demodulation. In Section IV, we formalize a general decision framework for PPM demodulation and derive expressions for the total probability of error under various strategies. Section V presents existing and newly proposed adaptive PPM receivers, beginning with optimized Direct Detection and Conditional Pulse Nulling (CPN), followed by an extended greedy receiver class incorporating joint displacement and squeezing as well as Dolinar-like waveform control. We then introduce the slicing Bayesian greedy receiver, which tracks full posterior distributions over codewords and uses slice-by-slice Bayesian updates to adapt measurements. Section VI presents numerical results under both noiseless and noisy conditions, comparing all receiver types. Finally, Section VII concludes with a discussion of implications, scalability, and future directions for structured quantum receivers.

\section{Detection statistics}

\begin{figure}[htbp]

\centering
\includegraphics[width=.6\linewidth]{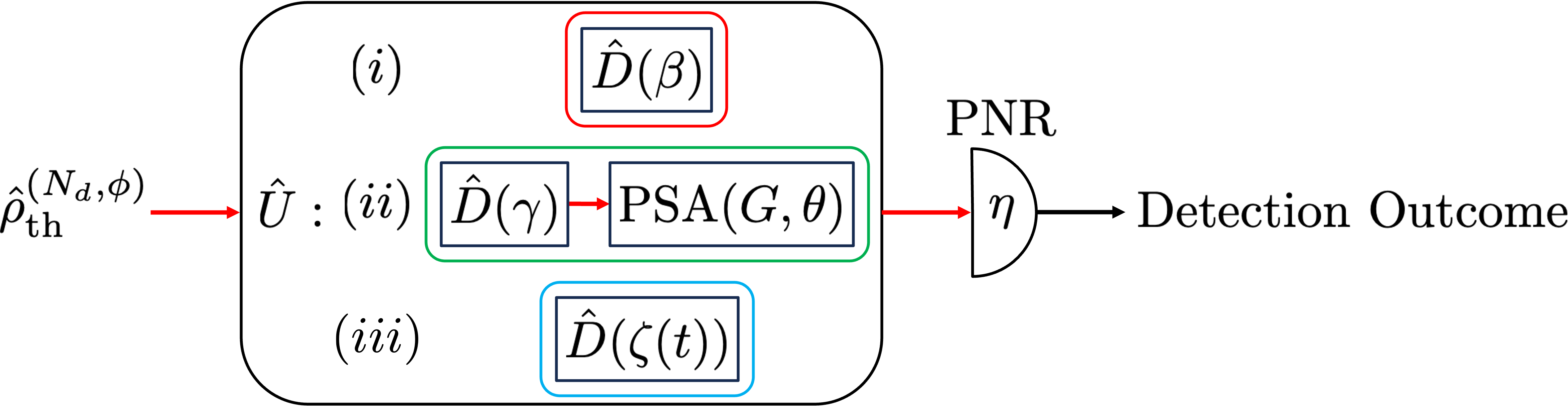}
\caption{Schematic of a structured optical receiver employing a unitary transformation followed by local detection.}
\label{fig:unitary}
\end{figure}

For our detection analysis, we consider the transmission of a general coherent state temporal waveform $\hat{\rho}^{(\phi)}=\ket{\phi}\bra{\phi}$, where $\ket{\phi}$ is the canonical coherent state of the harmonic oscillator with complex amplitude $\phi$, through a thermal classical additive noise channel, which yields a displaced thermal state $\hat{\rho}_{\text{th}}^{(N_d,\phi)}$, where $N_d$ is the mean thermal photon number. This channel models thermal fluctuations in the absence of loss and is formally defined as a Gaussian channel that adds classical noise without altering the signal amplitude. It is characterized by the transformation of the input displacement vector $d_{\text{in}}$ and the covariance matrix $V_{\text{in}}$ according to $d_{\text{out}}=X d_{\text{in}}$ and $V_{out}=X V_{in} X^{\top}+Y$, where $X=\mathbb{I}$ and $Y=\text{diag}(N_d,N_d)$. Here, $N_d\geq0$ represents the variance of the added classical Gaussian noise.

Upon reception of the displaced thermal state, one of three unitary operations $\hat{U}$ is applied, as illustrated in Fig.~\ref{fig:unitary}. Operation ($i$) corresponds to a displacement $\hat{D}(\beta)$, which coherently shifts the amplitude of the input state: $\hat{D}(\beta)\ket{\phi}=\ket{\phi+\beta}$. This is typically implemented using a highly transmissive beamsplitter and a strong local oscillator \cite{Guha_CPN_expt}. Operation ($ii$) applies a displacement $\hat{D}(\gamma)$, followed by a single-mode squeezing operations $\hat{S}(r,\theta)$, often referred to as a phase sensitive amplifier (PSA). Such an operations amplifies one quadrature of the field (aligned with the coherent state's phase when $\theta=0$) while de-amplifying the orthogonal quadrature. In this sense it acts as a phase sensitive linear amplifier, sometimes called 'noiseless' because the amplified quadrature does not acquire extra noise beyond vacuum fluctuations, in contrast to phase-insensitive amplification. Operation ($iii$) implements a time-varying displacement waveform $\hat{D}(\zeta(t))$.

After transmission through the thermal additive noise channel, the displaced thermal state is measured using a photon-number-resolving (PNR) receiver, which can resolve multiple independent photon detection events. Only relevant for operations ($i$) and ($ii$), a PNR receiver or Single Photon Detector (SPD) with sub-unity quantum efficiency $\eta\in(0,1]$, the probability of registering no detection events\footnote{{Eq,~\eqref{eqn:squeezed_noclick}} can be obtained by starting with a coherent state that propagates through the displacement thermal noise chancel. Upon arrival, a displacement and then the squeezing operation are applied. The state then passes through a lossy channel with efficiency $\eta$ before detection. The expression follows by writing the transformed displacement vector and covariance matrix, computing the resulting Wigner function, and integrating it against the Wigner function of the vacuum state to evaluate the no-click probability.} ("no-click" event) is given by
\begin{equation}
\label{eqn:squeezed_noclick}
p_{\phi,\Gamma}= \frac{\exp\left[ 
    -\frac{\eta\, (\sqrt{G} + \sqrt{G - 1})^2 \bar{N}}{1 + \eta \left( (1 + 2N_d)\sqrt{G - 1}(\sqrt{G - 1} + \sqrt{G}) + N_d \right)} 
\right]}{\sqrt{(2G - 1 + (1 - G + N_d)\eta)^2 + (G - 1)G(\eta - 2)^2}},
\end{equation}

where the first and second index is the transmitted state and operations applied, respectively, $\bar{N}=|\phi+\Gamma|^2$ is the mean photon number sum of the received temporal waveform $\phi$ and the applied displacement $\Gamma\in\{\beta,\gamma\}$ corresponding to the unitary operations. The click outcome is $\bar{p}_{\phi,\Gamma}=1-p_{\phi,\Gamma}$ where $p_{\phi,0}$ indicate no operation performed, e.i. $\bar{N}=|\phi|^2$. For operations ($i$), the phase-sensitive amplifier (PSA) gain is $G=1$, while for ($ii$) we assume squeezing with gain $G>1$ and phase $\theta=0$. Operation ($iii$), which involves a time-varying displacement strategy, will be discussed in detail in Section \ref{sec:dolinar}.
\section{Two Coherent-State Discrimination Strategies}
PPM sequential slot hypothesis update strategies can be fundamentally reduced to a series of binary discrimination problem between two coherent states: the vacuum state $\ket{0}$ and a nonzero coherent state $\ket{\alpha}$, here without loss of generality $\alpha\in\mathbb{R}$. Therefore, to fully characterize receiver designs for PPM demodulation, it is instructive to first review key strategies developed for binary coherent state discrimination. These include the Kennedy receiver, the optimally displaced Kennedy receiver, the optimally displaced and squeezed receiver, and the Dolinar receiver. Each of these represent a different trade-off between implementation complexity and proximity to the quantum-optimal measurement, and they serve as a foundational building blocks for slot-wise detection in PPM. The Helstrom limit for discriminating between the two state alphabet $\{\ket{0},\ket{\alpha}\}$ is given by
\begin{equation}
P_{\text{e},\min} = \frac{1}{2} \left(1 - \sqrt{1 - 4\xi(1 - \xi)e^{-N}} \right),
\label{eqn:helstrom}
\end{equation}
where $\xi$ and $1-\xi$ is the prior probabilities of transmitting $\ket{0}$ and $\ket{\alpha}$, respectively, and $N=|\alpha|^2$ is the mean photon number. Demonstrating $P_{\text{e,min}}\sim e ^{-N}$ in the large mean photon number $N\gg1$.
\subsection{Displacement and Squeezing Strategy}
Formally, the Kennedy receiver \cite{Kennedy1973} provides a near-optimal strategy for discriminating the Binary Phase Shift Key (BPSK) alphabet $\{\ket{\alpha},\ket{-\alpha}\}$.It operates by displacing one of the states to vacuum using $\hat{D}(\alpha)$ or $\hat{D}(-\alpha)$, transforming the alphabet into $\{\ket{0},\ket{\pm2\alpha}\}$, and then performing SPD. A hypothesis is made based on whether or not a photon is detected ("click" versus "no-click"). For the codeword alphabet $\{\ket{0},\ket{\alpha}\}$, a strategy analogous to the Kennedy receiver involves directly measuring the incoming pulse using a SPF, with a hypothesis formed based on whether a photon is detected. This approach---commonly referred to as Direct Detection (DD)---requires no displacement or interference and serves as a baseline receiver in many optical communication settings and achieving the error probability of $\tfrac{1}{2}e^{-N}$. The Kennedy receiver can be improved by implementing either operation ($i$) or ($ii$) from Fig \ref{fig:unitary}. In the operation ($i$), the displacement $\hat{D}(\beta)$ is optimized to minimize the probability of error, while in operation ($ii$), both the displacement $\hat{D}(\gamma)$ and the PSA gain $G$ are jointly optimized, as demonstrated in \cite{takeoka2008}. For both cases, DD serves as an upper bound on the probability of error, although these receiver do not achieve the Helstrom limit for the binary coherent-state alphabet.

\subsection{Dolinar Receiver}
\label{sec:dolinar}
The Dolinar receiver (DR)~\cite{dolinar1976,Dolinar_2011} provides a quantum-optimal strategy for binary coherent-state discrimination. Given two time-dependent coherent states $\{\psi_0(t), \psi_1(t)\}$ of duration $T$ and with inner product $\braket{\psi_0 | \psi_1} = e^{-N_0}$, where $N_0 = \int_0^T |\psi_0(t) - \psi_1(t)|^2 \, dt$, the DR applies a real-time feedback-based displacement $u_z(t)$ based on an internal hypothesis $z \in \{0,1\}$. The displaced signal is measured by an ideal single-photon detector with instantaneous rate
\begin{equation}
\lambda(t) = |\psi_j(t) + u_z(t)|^2,
\end{equation}
where $j\in\{0,1\}$ denotes the index of the actual transmitted state. Upon each photon detection event, the receiver instantaneously switches the current hypothesis ($z \to 1 - z$) and updates the applied displacement accordingly. The hypothesis held at time $t = T$ is declared as the final hypothesis. This decision rule can be equivalently described as declaring a hypothesis based on whether a total number of detection event (clicks) is even or odd.

If the receiver operates under an assumed prior $\xi_1$, which may differ from the true prior $\xi$, one can derive the time varying waveforms $ u_z(t)$, such that the conditional probabilities of detecting an odd number of photons are
\begin{align}
\begin{split}
P_{0}^{\text{odd}} &= \frac{1}{2} \left(1 - \frac{1 - 2\xi_1 e^{-N_d}}{\sqrt{1 - 4\xi_1(1 - \xi_1)e^{-N_d}}} \right), \\
P_{1}^{\text{odd}} &= \frac{1}{2} \left(1 + \frac{1 - 2(1 - \xi_1)e^{-N_d}}{\sqrt{1 - 4\xi_1(1 - \xi_1)e^{-N_d}}} \right),
\label{eqn:dolinar_noclick}
\end{split}
\end{align}
with $P_{j}^{\text{even}} = 1 - P_{j}^{\text{odd}}$. The average error probability with true prior $\xi$ is
\begin{equation}
P_{\text{e}} = \xi \, P_{0}^{\text{even}} + (1 - \xi) \, P_{1}^{\text{odd}}.
\label{eqn:dolinar_pe}
\end{equation}
When considering our alphabet $\{\ket{0},\ket{\alpha}\}$ and $\xi_1 = \xi$, this expression achieves the Helstrom bound eqn. (\ref{eqn:helstrom}). 
\section{PPM Receiver Decision Framework and Error Probability}

For any discrete-time demodulation strategy, detection proceeds slot by slot, generating a binary sequence of measurement outcomes. Let $\vec{c}_k = [c_1 c_2\dots c_k]$ denote the sequence of click outcomes observed up to and including time slot $k$, where each $c_j \in \{0, 1\}$ indicates a no-click ($0$) or a click ($1$) event. Based on this observed sequence, the receiver maintains or updates an internal hypothesis $h_k \in \{1, \dots, M\}$ representing its current estimate of the transmitted codeword. The decision rule for updating or selecting this hypothesis varies across receiver architectures and may range from greedy strategies to fully Bayesian inference.

For our error analysis, let $i \in \{1, \dots, M\}$ denote the true transmitted codeword. We define the probability of observing a no-click in the $k$-th time slot, conditioned on the prior measurement outcomes $\vec{c}_{k-1}$, as
\begin{equation}
P_k[0 \mid i, \vec{c}_{k-1}],
\end{equation}
with the corresponding click probability given by $P_k[1 \mid i, \vec{c}_{k-1}] = 1 - P_k[0 \mid i, \vec{c}_{k-1}]$. These probabilities depend on the transmitted codeword $i$, which determines whether a pulse ($i = k$) or vacuum ($i \neq k$) is present in slot $k$, as well as on the receiver’s measurement configuration at that slot. This configuration may adapt based on the prior outcomes $\vec{c}_{k-1}$ and can include displacement, squeezing, or a time-varying waveforms. The no-click probability $P_k[0 \mid i, \vec{c}_{k-1}]$ can be evaluated using the general displacement-and-squeezing model in Eq.~\eqref{eqn:squeezed_noclick}, or using the lossless Dolinar receiver in Eq.~\eqref{eqn:dolinar_noclick}. The probability of observing the click sequence $\vec{c}_k$ given that the $i$-th codeword was transmitted can be computed recursively as
\begin{equation}
P[\vec{c}_k \mid i] = P_k[c_k \mid i, \vec{c}_{k-1}] P[\vec{c}_{k-1} \mid i].
\end{equation}

Given the observed click sequence $\vec{c}_k$, the posterior probability that codeword $i$ was transmitted is computed via Bayes’ rule:
\begin{equation}
P[i \mid \vec{c}_k] = \frac{P[\vec{c}_k \mid i] \cdot P[i]}{\sum_{j=1}^M P[\vec{c}_k \mid j] \cdot P[j]},
\label{eqn:bayespiror}
\end{equation}
where $P[i]$ is the prior probability of transmitting codeword $i$.

The receiver selects a hypothesis $h_k \in \{1, \dots, M\}$ based on the observed click outcome sequence $\vec{c}_k$. This hypothesis represents the receiver’s estimate of the transmitted codeword after measuring all $k$ slots. Its selection may depend on the full click history and can be defined according to a variety of decision strategies.

The total probability of correct decision, denoted $P_{\text{c}}$, is obtained by averaging over all possible transmitted codewords and measurement outcome sequences:
\begin{equation}
\label{eqn:Pc}
P_{\text{c}} = \sum_{i=1}^M P[i] \sum_{\vec{c}_M} P[\vec{c}_M \mid i]  P[h_M = i \mid \vec{c}_M],
\end{equation}
where $P[h_k = i \mid \vec{c}_k]$ denotes the probability that the receiver selects hypothesis $i$ given the observed click sequence $\vec{c}_k$. This quantity depends on the specific decision strategy employed by the receiver, which may be deterministic or probabilistic. The total probability of error is then given by $P_{\text{e}} = 1 - P_{\text{c}}$.

\subsection{Direct Detection, Conditional Pulse Nulling, and other Adaptive Approaches}
\label{sec:DD,CPN,Adaptive}
The baseline for PPM demodulation is DD, where each slot is detected using a SPD, any slot that produced a click will be randomly uniformly be selected as the final hypothesis. This approach typically does not apply any operation from Fig. \ref{fig:unitary}, for our analysis we will be consider the case where we apply operation ($i$) from Fig. \ref{fig:unitary} for analysis, although any operation ($i$)-($iii$) can be used. The probability of error for DD of PPM can be derived using eqn. (\ref{eqn:Pc}), and is present in \cite{Lukanowski:24} as
\begin{equation}
    \label{eqn:DD_PPM}
    P_{\mathrm{e,DD}} = \frac{(M - p_{\Gamma} q_{\Gamma}^{M-1}) \bar{q}_{\Gamma} - \bar{p}_{\Gamma} (1 - q_{\Gamma}^M)}{M \bar{q}_{\Gamma}}
\end{equation}
while adopting notation $p_\Gamma=p_{\alpha,\Gamma}$ and $q_\Gamma=p_{0,\Gamma}$. In the high-photon-number regime ($N \gg 1$), and for fixed $M$, the error probabilities scale as \( P_{\textrm{e,DD}} \sim e^{-N} \) for DD which does not achieve the Helstrom scaling \( P_{\mathrm{e,min}} \sim e^{-2N} \).

The Conditional Pulse Nulling (CPN) receiver originally proposed by Dolinar \cite{dolinar_CPN}, applies a adaptive displacement to the $M$ PPM time slots based on intermediate detection outcomes. The receiver begins with an internal hypothesis $z$, assuming the pulse is in the first slot. It then applies a displacement $\hat{D}(-\alpha)$ to that slot---effectively attempting to null the assumed present coherent state---and performs SPD. If a click is registered the hypothesis is incremented $(z\rightarrow z+1)$, and the nulling procedure is repeated fro the next slot. If no click is observed, the receiver interprets this as evidence that the pulse has been successfully nulled and proceed to DD the remaining slots $k+1$ to $M$. If any of those subsequent slots yield a click, the internal hypothesis is updated to the corresponding slot. If no additional clicks occur, the receiver retains the current hypothesis $z$ as the final hypothesis. The nulling phase of the CPN receiver has been enhanced by implementing a fixed optimal operations, either displacement-only ($i$) or displacement followed by squeezing ($ii$) \cite{Guha10022011}. Further improvement is possible by employing the DR ($iii$), which updates the displacement waveform based on prior detection outcome. Considering these case, the probability of error which can be derived from eqn. (\ref{eqn:Pc}) and presented in \cite{Lukanowski:24} as
\begin{equation}
    P_{\mathrm{e,CPN}} = \frac{
    (\bar{p}_0 - M \bar{q}_0)(q_0 - \bar{q}_{\Gamma})
    + \bar{q}_0 \bar{q}_{\Gamma}^{M-1} (\bar{p}_{\Gamma} q_0 - p_0 \bar{q}_{\Gamma})
    + q_0^M (p_{\Gamma} \bar{q}_0 - \bar{p}_0 q_{\Gamma})
}{M\bar{q}_0(q_{\bar{\Gamma}} - q_0)}  
\end{equation}
The CPN receiver achieves Hlestrom-like scaling in its probability of error, with $P_{\mathrm{e,CPN}} \sim e^{-2N}$ when the number of PPM slots satisfies $Me^{-N}\ll1$.

Beyond the CPN receiver, more recent approaches have explored fully adaptive receiver architectures \cite{adaptive} that numerically optimize the operations ($i$)-($iii$) at each time slot on prior detection outcomes. These strategies construct and a traverse a binary decision tree of depth $M$, where each node corresponds to a section history vector $\vec{c}_k$ and prescribes an optimal operation for each slot. Upon receiving a measurement outcome, the receiver follows the corresponding branch, updates its hypothesis, and applies the next displacement accordingly, This allows for dynamic, history-dependent adaptation of the measurement strategy, unlike CPN, which uses a fixed update rule. While this framework has demonstrated promising performance for small $M$, the exponential growth of the tree with increasing $M$ imposes a severe memory and computational bottleneck.

\begin{algorithm}
\caption{Greedy Receiver}
\begin{algorithmic}[1]
\State $h_1 \gets 1$ \Comment{Initialize hypothesis to slot 1}
\State $\mathcal{M}_1 \gets$ initial measurement setting (e.g., displacement and squeezing, or waveform)

\State Apply $\mathcal{M}_1$ to slot 1 and detect to obtain $c_1 \in \{0, 1\}$
\State $r_1 \gets$ initial revision ratio based on $c_1$

\For{$k = 2$ to $M$} \Comment{Iterate through remaining slots}
    \State Evaluate two update strategies:
    \State \hspace{1em} Option A: update hypothesis if $c_k = 1$
    \State \hspace{1em} Option B: update hypothesis if $c_k = 0$
    \State For each option, compute optimal measurement setting $\mathcal{M}_k$ maximizing $P_{\mathrm{c}}$
    \State Select option (A or B) with higher $P_{\mathrm{c}}$ \Comment{Greedy choice}
    \State Apply selected measurement $\mathcal{M}_k$ to slot $k$
    \State Detect slot to obtain $c_k \in \{0, 1\}$
    \If{update condition for selected option is satisfied}
        \State $h_k \gets k$ \Comment{Update hypothesis to current slot}
        \State Update revision ratio $r_k$ based on $c_k$
    \Else
        \State $h_k \gets h_{k-1}$ \Comment{Retain previous hypothesis revision ratio}
        \State $r_k \gets r_{k-1}$
    \EndIf
    
\EndFor
\State \textbf{return} $h_M$ \Comment{Final hypothesis after $M$ slots}
\end{algorithmic}
\label{alg:greedy}
\end{algorithm}

\subsection{Greedy Receiver Strategy}
Łukanowski's greedy receiver~\cite{Lukanowski:24} maintains a single internal parameter called the \textit{revision ratio}, which quantifies how favorable it is to revise the current hypothesis based on the observed outcome and the measurement setting. At slot $k$, the receiver applies a measurement configuration $\mathcal{M}_k$ (e.i. operations ($i$)-($iii$) from Fig. \ref{fig:unitary}, displacement,  joint displacement and squeezing, or waveform). Based on the click outcome $c_k$, the revision ratio is defined as:
\begin{equation}
r_k =
\begin{cases}
\displaystyle\frac{P[1 \mid h_k \neq i, \mathcal{M}_{k}]}{P[1 \mid h_k = i, \mathcal{M}_{k}]} & \text{if } c_{k} = 1, \\[10pt]
\displaystyle\frac{P[0 \mid h_k \neq i, \mathcal{M}_{k}]}{P[0 \mid h_k = i, \mathcal{M}_{k}]} & \text{if } c_{k} = 0.
\end{cases}
\end{equation}

At each subsequent slot $k$, the receiver considers two possible update strategies:
\begin{itemize}
    \item \textbf{Option A:} Update the hypothesis and revision ratio if a no-click is observed ($c_k = 0$)
    \item \textbf{Option B:} Update the hypothesis revision ratio if a click is observed ($c_k = 1$)
\end{itemize}

For each option, the receiver selects a measurement setting $\mathcal{M}_k$ that maximizes the probability of a correct final decision. Let $p_{\mathcal{A}}$ and $p_{\mathcal{B}}$ denote the resulting success probabilities:
\begin{align}
\begin{split}
p_{\mathcal{A}} &= P[0 \mid k \neq i, \mathcal{M}_{k}] + r_k  P[1 \mid k = i, \mathcal{M}_{k}], \\
p_{\mathcal{B}} &= r_k  P[0 \mid k = i, \mathcal{M}_{k}] + P[1 \mid k \neq i, \mathcal{M}_{k}]).
\label{eqn:optAoptB}
\end{split}
\end{align}
The receiver selects the option achieving $\max\{p_{\mathcal{A}}, p_{\mathcal{B}}\}$ and applies the corresponding measurement to slot $k$. If the update condition is met, the hypothesis is set to $h_k \gets k$, and the revision ratio $r_k$ is updated accordingly. The initial measurement setting $\mathcal{M}_1$ is chosen to minimize the overall probability of error. The full process is detailed in Algorithm \ref{alg:greedy}. The greedy receiver stands out for its strong performance under both ideal and noisy conditions, despite relying on locally optimal update rules. Its robustness to mode mismatch and noise, efficient implementation via lookup tables, and extensibility to richer measurement schemes make it highly practical and adaptable architecture. It shows potential for broader applications beyond PPM, including Quantum Phase Shift Keying (QPSK) demodulation, and can incorporate erasure-handling mechanism to support capacity based evaluations.

\begin{algorithm}
\caption{Slicing Receiver}
\begin{algorithmic}[1]
\State Initialize priors $P_0[i] \gets 1/M$ for all $i \in \{1, \dots, M\}$
\State $k \gets 1$ \Comment{Slice index from 1 to $Mn$}
\While{$k \leq Mn$}
    \For{$i = 1$ to $M$} \Comment{Iterate through each PPM slot}
        \If{$r_k < 0$}
            \State Apply Option A: update hypothesis and revision ratio
        \Else
            \State Apply Option B: update revision ratio and retain or update hypothesis
        \EndIf
        \State For selected option, compute optimal measurement setting $\mathcal{M}_k$
        \State Apply selected measurement $\mathcal{M}_k$ to slice $k$
        \State Detect slice to obtain $c_k \in \{0, 1\}$
        \State Update prior for hypotheses using Bayes' rule:
            \State \hspace{1em} $P_{k}[i] \propto P[c_k \mid i, \mathcal{M}_k] \cdot P_{k-1}[i]$

    \EndFor
        \State $k \gets k + 1$
\EndWhile
\State $h \gets \arg\max_i P_{Mn}[i]$ \Comment{Final hypothesis after all slices}
\State \textbf{return} $h$
\end{algorithmic}
\label{alg:slicing}
\end{algorithm}
\subsection{Bayesian Slicing Greedy Receiver}

Each PPM time slot can be optically sliced into $n$ components using a cascade of $n-1$ beamsplitters. The beamsplitters are configured with sequential trasmissivites $\{\tfrac{n-1}{n},\tfrac{n-2}{n-1},
\dots,\tfrac{1}{2}\}$. At each stage, the transmitted (upper) port feeds into the next beamsplitter in the cascade, while the reflected (lower) port corresponds to one of the $n$ optical slices. When the displaced thermal state $\hat{\rho}_{\text{th}}^{(N_d,\phi)}$ is feed into this cascade of beam splitters it yeilds $n$ identical slices $\hat{\rho}_{\text{th}}^{(N_d/n,\phi/\sqrt{n})}$, each with reduced mean thermal photon number $\tfrac{N_d}{n}$ and scaled displacement amplitude $\tfrac{\phi}{\sqrt{n}}$.

The resulting PPM codewords are described as tensor products of displaced thermal states across all $Mn$ slices:
\begin{equation}
\hat{\rho}_i = \bigotimes_{j=1}^M \bigotimes_{k=1}^n \hat{\rho}_{j,k},
\end{equation}
where the state in slice $(j,k)$ is given by
\begin{equation}
 \hat{\rho}_{j,k} = 
\begin{cases}
\hat{\rho}_{\text{th}}^{(N_d/n,\alpha/\sqrt{n})} & \text{if } j = i, \\
\hat{\rho}_{\text{th}}^{(N_d/n,0)} & \text{otherwise}.
\end{cases}
\end{equation}

Inspired by the Greedy Receiver framework and by prior result including slicing for demodulating PPM \cite{bia}, the slicing receiver maintains a running prior distribution over all codeword hypotheses, updated after each slice based on the observed click outcome \(c_k\), using Bayes’ rule as given in Eq.~\eqref{eqn:bayespiror}. At each slice \(k\), the receiver considers two update strategies—Option A (update on no-click) and Option B (update on click)—and uses a local optimization procedure to maximize the probability of a correct decision. In contrast to the Greedy Receiver the Slicing Greedy receiver will always choose option A when $r_k\leq1$ and and option B when $r_k>1$. 

To guide this decision, the receiver computes a \textit{revision ratio} defined as the likelihood ratio between the current most probable hypothesis and the second-best alternative:
\begin{equation}
    r_k = \frac{P[k \mid \vec{c}_{k-1}]}{\max\limits_{j \in \{1,\dots,M\} \setminus \{k\}}\left[ P[j \mid \vec{c}_{k-1}]\right]}.
\end{equation}

The measurement setting $\mathcal{M}_k$ in the slice \(k\) is chosen to maximize the resulting success probabilities, as described in Eq.~\eqref{eqn:optAoptB}. After all slices have been measured, the final hypothesis is selected as:
\begin{equation}
h_k = \arg\max_{j \in \{1, \dots, M\}} \left[P[j \mid \vec{c}_k]\right].
\end{equation}
The full process is detailed in Algorithm \ref{alg:slicing}. 

\begin{figure}[H]
    \centering
    \includegraphics[width=0.7\textwidth]{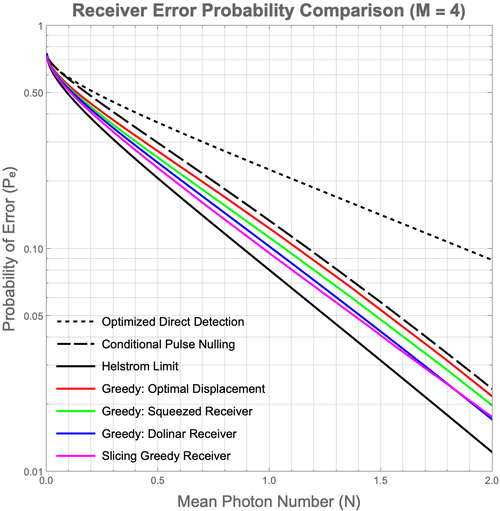}
    \caption{Receiver error probability for 4-PPM comparison under noiseless conditions $N_d=0$ in linear-log scale.}
    \label{fig:4PPM_noiseless}
\end{figure}
\section{Results}
Figure~\ref{fig:4PPM_noiseless}, shows the probability of error for 4-PPM under noiseless conditions ($N_d=0$) as a function of mean photon number $N\in[0,2]$, plotted on a semi-logarithmic (linear-log) scale. The solid black curve represents the Helstrom limit, serving as the theoretical benchmark for minimum achievable error probability. The performance of DD and CPN receivers are shown as short-dashed and long-dashed black curves, respectively. In both case operation the ($i$) from Fig. \ref{fig:unitary} was used, where the displacement $\hat{D}(\beta))$ was optimized to minimize the probability of error. Three variants of the greedy receiver were evaluated, each corresponding to a different measurement operations: ($i$) optimal displacement $\hat{D}(\beta))$ (Red), ($ii$) jointly optimized displacement $\hat{D}(\gamma))$ and PSA gain $G$ (Green), and ($iii$) time-varying waveform $\hat{D}(\zeta(t))$ DR (Blue). Each successive configuration from ($i$) to ($iii$) achieves improved performance, highlighting the benefit of increase measurement adaptability. Finally, the slicing greedy receiver, shown as the magenta curve, was evaluated using Monte Carlo simulation with the number of optical slices set as $n=1000N$, scaling with the mean photon number $N$. For this receiver, operation $(i)$ from Fig.~\ref{fig:unitary} was applied at each slice. A total of $10^5$ trials were conducted and the probability of error was evaluated after each slice $n$. This receiver achieve the lowest error probabilities in the low-mean-number regime and outperforms all other strategies for $N\lesssim1.8$, making it the most effective demodulation strategy in this regime.

\begin{figure}[H]
    \centering
    \includegraphics[width=0.7\textwidth]{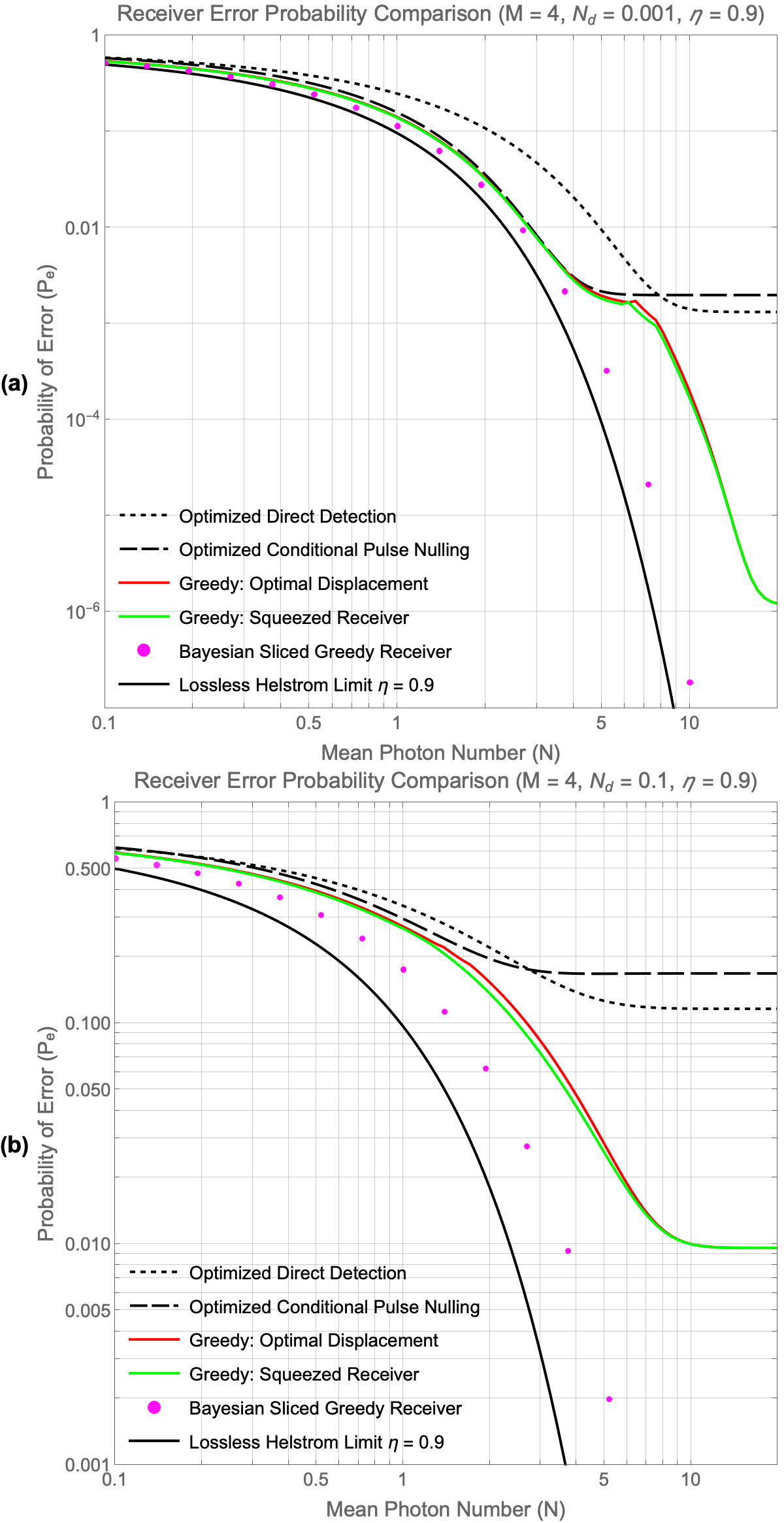}
    \caption{Receiver error probability for 4-PPM comparison under noisy conditions $N_d=0.001$ for (a) and $N_d=0.1$ for (b) both with sub-unity quantum efficiency $\eta=0.9$ in log-log scale.}
    \label{fig:4PPM_noise}
\end{figure}

Figure \ref{fig:4PPM_noise} shows the probability of error for 4-PPM under noisy conditions, plotted on a log-log scale as a function of mean photon number $N\in[0.1,20]$. Part (a) corresponds to a moderate noise level with thermal mean photon number $N_d=0.001$, while part (b) presents a more sever noise case with $N_d=0.1$. In both case, detection is performed with sub-unity quantum efficiency $\eta=0.9$. The performance of DD and CPN receiver are shown in each part. For both, the optimal ($i$) operational was applied. Two variants of the greedy receiver are included corresponding to operations ($i$---red) and ($ii$---Green). The optimally displaced and squeezed greedy variant does not perform much better than the optimally displaced version. Although, both greedy receivers perform comparably to the CPN receiver, with only marginal gains. However, in the high-photon-number regime, the greedy receiver exhibit significantly lower error probabilities, with a clearly reduced noise floor. For example, under moderate noise ($N_d=0.001$), the greedy receiver achieves error probabilities approaching $10^{-6}$ while under stronger noise ($N_d=0.1$), the floor is approximately $10^{-3}$, These improvements highlight the advantage of adaptive measurement selection when sufficient signal power is available. The slicing greedy receiver, shown as magenta dots, demonstrates the best overall performance in both noise scenarios. 
For this receiver, operation~(i) is applied at each slice and each magenta data point was obtained by Monte Carlo simulation with $n=1000$ slices and $10^5$ trials per point. 
Across all mean photon numbers, it outperforms both CPN and standard greedy receivers. 
Notably, the slicing Bayesian strategy exhibits no observable noise floor, even under thermal background conditions, in contrast to all other receivers considered. 
This phenomenon is explained in detail in Appendix~A, where we show that repeated Bayesian updates allow the posterior to continue concentrating on the correct slot, preventing the plateau behavior that arises in conventional direct detection.

\section{Conclusion}

In this work, we have extended the framework of adaptive quantum receivers for Pulse Position Modulation (PPM) by incorporating both phase-sensitive amplification and fine-grained optical slicing into the greedy decision-making process. We first introduced squeezed-state operations into the greedy receiver architecture, enabling an additional degree of control over the detection statistics through the optimization of the displacement and the phase-sensitive amplifier (PSA) gain. This enhancement provides measurable gains over displacement-only strategies, particularly in the low-photon-number regime, and further narrows the gap to the Helstrom bound using only local operations. 

Most significantly, we proposed and analyzed the slicing greedy receiver, a hybrid Bayesian greedy approach that slices each optical pulse into multiple subcomponents and updates the hypothesis distribution after every detection event. This design leverages the same revision-based philosophy as the standard greedy receiver but allows for far finer information extraction. Our simulations show that the slicing greedy receiver outperforms all other strategies across the full range of mean photon numbers, and—crucially—does not exhibit an apparent error floor under thermal background noise. This strongly suggests that repeated Bayesian updating is responsible for eliminating the plateau behavior observed in conventional direct detection. At present, however, this ``noise-floor-free'' behavior is an empirical observation, and a full analytic treatment of the Bayesian direct-detection PPM slicing receiver remains open. Developing such an analytic description may provide deeper intuition into the role of Bayesian inference in approaching the Helstrom limit.

While the slicing receiver offers strong performance, its computational overhead, requiring storage and update of the full posterior distribution, makes it less scalable than the original greedy architecture, which operates with only a revision ratio and current hypothesis. This tradeoff between inference resolution and resource efficiency frames an important frontier for future receiver design. 

Further advances in PPM demodulation may come from analyzing receiver algorithms that improve achievable capacity beyond structured joint detection in the photon-starved regime. Additionally, adaptively optimizing $2^M\!-\!1$ measurement strategies for long codewords remains computationally prohibitive. A promising direction is to formulate receiver strategies in continuous time using time-varying waveforms, which may lead to both performance gains and tractable closed-form solutions.
\begin{backmatter}
\bmsection{Funding}
C.N.G. and L.B. acknowledge the award ARO W911NF-24-1-0080.

\bmsection{Acknowledgments}
The authors are grateful to Karol Łukanowski for fruitful discussions.

\bmsection{Disclosures}
The author declares no conflicts of interest.

\bmsection{Data availability}
The data that support the findings of this study are available from the corresponding author upon reasonable request.
\end{backmatter}

\clearpage
\bibliography{refs}
\clearpage
\appendix
\section{Bayesian Direct-Detection Slicing Receiver for \texorpdfstring{$M$}{M}-PPM}
\label{app:BayesDD}

We consider $M$-ary PPM with $M=4$ and unity quantum efficiency ($\eta=1$). The $i$th noiseless codeword is defined in Eq. \eqref{eqn:PPM}. Let $N$ denote the mean signal photons per symbol (x-axis range $N\in[0.1,30]$, log scale). Thermal background is denoted $N_d$. Click/no-click statistics follow the model already presented generally as Eq. \eqref{eqn:squeezed_noclick}, for the Bayesian Direct Detection (DD) slicing results here we simplify to the no-operations case (no displacement, no squeezing ) by setting $\Gamma=0$ and $G=1$ in Eq. \eqref{eqn:squeezed_noclick}.

Each of the $M$ time slots is partitioned into $n$ equal-durations slices, producing $Mn$ sequential binary observations (click, no-click), detected in series. No optical operations are applied, slicing changes only the amplitude of received mean photon number $N/n$ and mean thermal photon number $N_d/n$. After each slice, the posterior over the $M$ hypotheses is updated using the Bayesian rule already defined in the paper, see Eq. \eqref{eqn:bayespiror}. The final decision is the index with the largest posterior probability at the end of the $Mn$ observations.

Standard unsliced DD (one decision per slot) exhibits an apparent error floor at high $N\gg1$ in the presence of thermal background noise because sporadic off-slots clicks persist and dominate tie-breaking, see the closed-form DD expression and discussion in Section \ref{sec:DD,CPN,Adaptive}. In contrast, slicing provides many fine-grained observations per symbol. Each slice contributes a small amount of inference, repeatedly applying Bayes' rule Eq. \eqref{eqn:bayespiror} concentrates posterior mass on the true slot while background-driven off-slot clicks still exists, so $P_e$ continues to decrease with $N$ rather than plateauing. An intuition is the "two nearly identical coins" analogy: with enough flips (here, slices), the posterior separates the hypotheses even when individuals outcomes are noisy. This phenomenon is purely Bayesian-it arises without any physical enhancement (no squeezing, no displacement) and is observed empirically in Fig. \ref{fig:bayes_slicing_combined} below. The DD Bayesian error probabilities were estimated using $10^6$ Monte Carlo trials across the curve. Across five background noise levels $N_d\in\{10^{-3},10^{-2},10^{-1},10^{0},10^{1}\}$. For orientation we also include the lossless ($N_d=0$, $\eta=1$) direct-detection Eq. \eqref{eqn:DD_PPM} reference and the lossless Helstrom limit for $M=4$ PPM Eq. \eqref{eqn:helstrom_PPM},

\begin{figure}[t]
    \centering
    \includegraphics[width=0.62\linewidth]{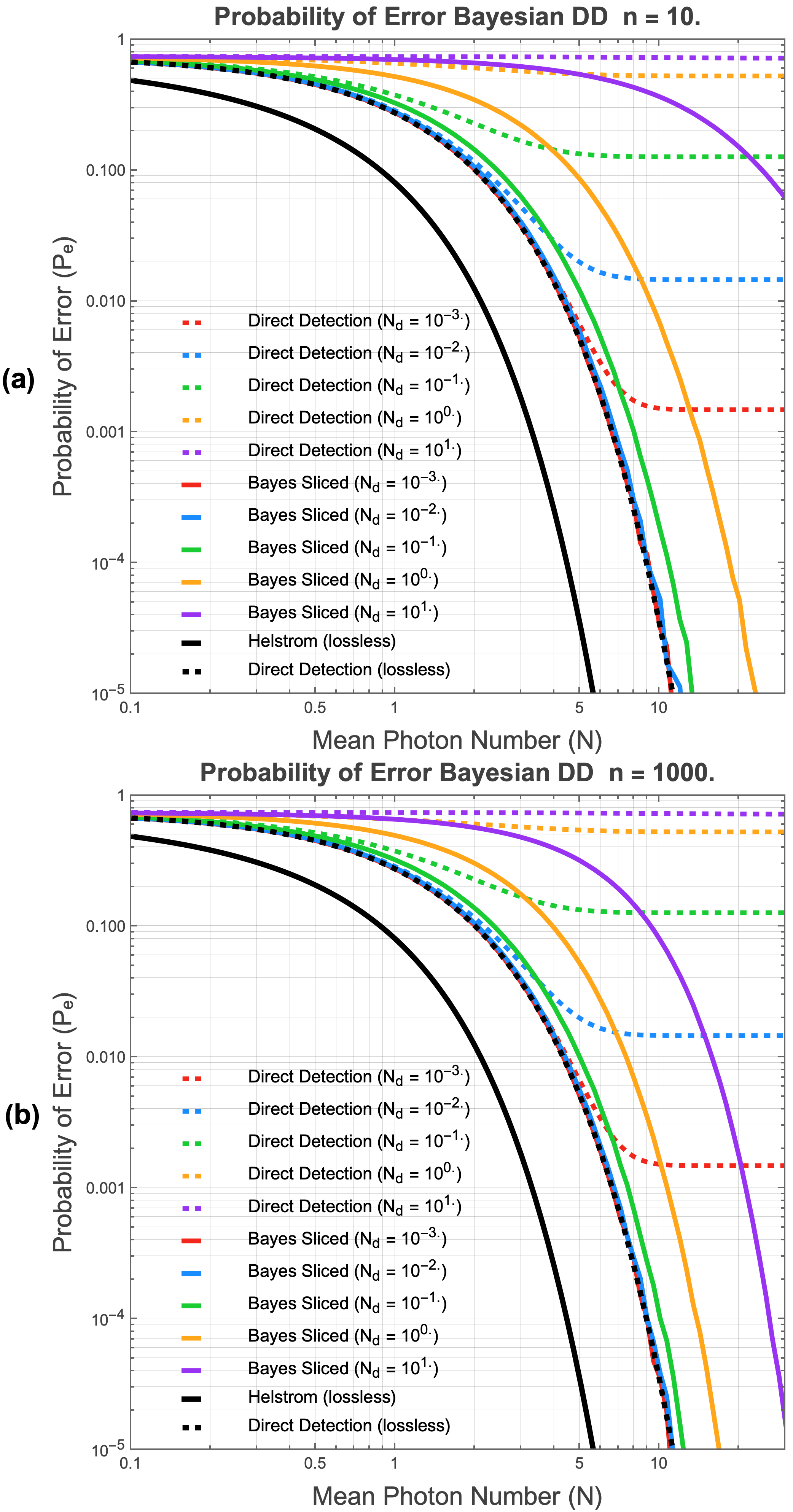}
    \caption{
    \textbf{Bayesian DD slicing receiver with $n{=}10$ (a) and $n{=}1000$ (b) slices per slot.}
      Error probability for $M{=}4$ PPM under unity efficiency ($\eta{=}1$) with background noise levels
      $N_d\in\{10^{-3},10^{-2},10^{-1},10^{0},10^{1}\}$.
      The x-axis is the mean photon number per slot $N\in[0.1,30]$ (log scale),
      the y-axis is $P_e\in[10^{-5},1]$ (log scale).
      Solid curves: Bayesian DD slicing using the click/no-click model from Eq. \eqref{eqn:squeezed_noclick}
      simplified to $\Gamma{=}0$, $G{=}1$ with sequential posterior updates via Eq. \eqref{eqn:bayespiror}.
      Dashed curves: standard unsliced DD baseline Eq. \eqref{eqn:DD_PPM},
      illustrating the apparent noise floor at high $N$.
      Reference lines: lossless Helstrom (Solid Black) for $M{=}4$ Eq. \eqref{eqn:helstrom_PPM} and
      lossless DD (Dashed Black) ($N_d{=}0$, $\eta{=}1$) Eq. \eqref{eqn:DD_PPM}.
      \textbf{(a)}~For $n{=}10$, slicing visibly suppresses the apparent floor relative to DD.  
      \textbf{(b)}~For $n{=}1000$, the suppression is even stronger, further lowering $P_e$ at fixed $N$ and confirming that repeated Bayesian updates alone can overcome background-induced plateaus. All DD Bayesian curves results are averaged over $10^6$ Monte Carlo trials.
      }
    \label{fig:bayes_slicing_combined}
\end{figure}

\end{document}